\documentclass[a4paper, twocolumn]{revtex4}
\usepackage[dvips]{graphicx} 
\usepackage{amsmath}
\usepackage{amsfonts}
\usepackage{amssymb}

\newcommand{\be}{\begin{equation}}
\newcommand{\ee}{\end{equation}}

\newcommand{\ef}[1]{\, #1}

\newcommand{\eval}[1]{\left\langle {#1} \right\rangle} 
\DeclareMathOperator*{\argmax}{argmax}
\DeclareMathOperator*{\argmin}{argmin}
\def\smfrac#1#2{{\textstyle\frac{#1}{#2}}}

\begin{document}

\title{Exact sampling of corrugated surfaces}

\author{Sergio Caracciolo}
\affiliation{Dip.~Fisica,
  Universit\`a degli Studi di Milano, and INFN, via G.~Celoria 16, 20133 Milano, Italy}
\author{Enrico Rinaldi}
\affiliation{Dip.~Fisica,
  Universit\`a degli Studi di Milano, and INFN, via G.~Celoria 16, 20133 Milano, Italy}
\author{Andrea Sportiello}
\affiliation{Dip.~Fisica,
  Universit\`a degli Studi di Milano, and INFN, via G.~Celoria 16, 20133 Milano, Italy}

\date{\today}

\begin{abstract}
\noindent
We discuss an algorithm for the exact sampling of vectors $\vec{v} \in [0,1]^N$ satisfying a set of pairwise difference
inequalities. Applications include the exact sampling of skew Young Tableaux, of configurations in the Bead Model, and of corrugated surfaces on a graph, that is random landscapes in which at each vertex corresponds a local maximum or minimum. As an example, we numerically evaluate with high-precision the number of corrugated surfaces on the square lattice. After an extrapolation to the thermodynamic limit, controlled by an exact formula, we put into evidence  a discrepancy with previous numerical results. 
\end{abstract}

\maketitle

\section{Introduction}
\label{sec.intro}

\noindent
Consider the following problem: given a grid $L \times L$, in how many
ways can we fill the boxes with the numbers from 1 to $L^2$ in such a
way that odd/even boxes are local maxima/minima? A typical allowed
configuration for $L=5$ is
\[
\setlength{\unitlength}{16pt}
\begin{picture}(6,5.7)(0,0.2)
\put(0,0.15){\includegraphics[scale=0.8, bb=0 0 120 120, clip=true]{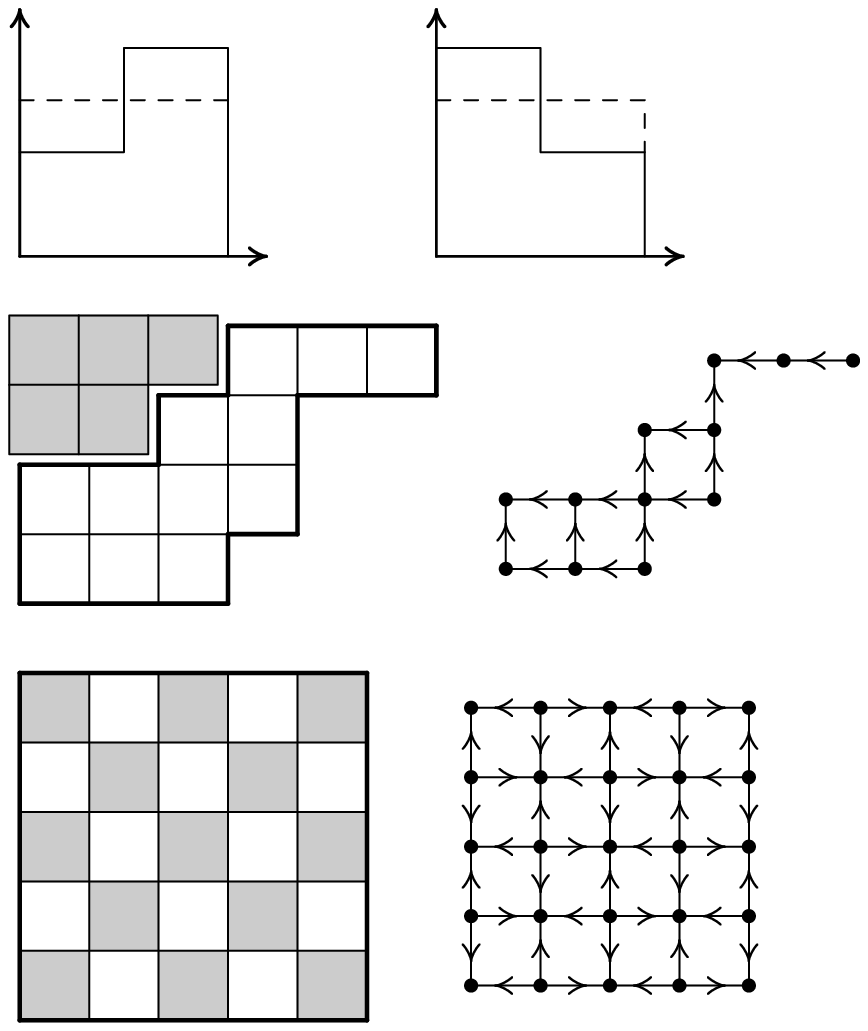}}
\put(1,1){\makebox[0pt][c]{2}}
\put(2,1){\makebox[0pt][c]{17}}
\put(3,1){\makebox[0pt][c]{1}}
\put(4,1){\makebox[0pt][c]{25}}
\put(5,1){\makebox[0pt][c]{19}}
\put(1,2){\makebox[0pt][c]{23}}
\put(2,2){\makebox[0pt][c]{5}}
\put(3,2){\makebox[0pt][c]{13}}
\put(4,2){\makebox[0pt][c]{4}}
\put(5,2){\makebox[0pt][c]{22}}
\put(1,3){\makebox[0pt][c]{9}}
\put(2,3){\makebox[0pt][c]{15}}
\put(3,3){\makebox[0pt][c]{11}}
\put(4,3){\makebox[0pt][c]{12}}
\put(5,3){\makebox[0pt][c]{7}}
\put(1,4){\makebox[0pt][c]{21}}
\put(2,4){\makebox[0pt][c]{8}}
\put(3,4){\makebox[0pt][c]{14}}
\put(4,4){\makebox[0pt][c]{3}}
\put(5,4){\makebox[0pt][c]{24}}
\put(1,5){\makebox[0pt][c]{6}}
\put(2,5){\makebox[0pt][c]{20}}
\put(3,5){\makebox[0pt][c]{10}}
\put(4,5){\makebox[0pt][c]{18}}
\put(5,5){\makebox[0pt][c]{16}}
\end{picture}
\]
A combinatorial more general setting is the following. Consider a
direct acyclic graph $G=(V, \vec{E})$, with $N$ vertices: how many
one-to-one maps $\sigma: V(G) \to \{ 1, \ldots N \}$ exist such that,
for each oriented edge $(ij) \in \vec{E}(G)$, $\sigma(i)>\sigma(j)$?
The case of the square grid pertinent to Corrugated Surfaces
corresponds to the graph
\[
\includegraphics[scale=0.8, bb=130 10 230 110, clip=true]{fig_grid1.eps}
\]
This wider framework includes, among other things, the counting 
of Standard Young Tableaux and Standard Skew Young 
Tableaux \cite{aitken, stanley}, as in the example below with the
skew tableau $(6,4,4,3) \smallsetminus (3,2)$:
\[
\setlength{\unitlength}{16pt}
\begin{picture}(12.3,4.7)(0,0.5)
\put(0,0.15){\includegraphics[scale=0.8, bb=0 120 260 220, clip=true]{fig_grid1.eps}}
\put(1,1){\makebox[0pt][c]{7}}
\put(2,1){\makebox[0pt][c]{9}}
\put(3,1){\makebox[0pt][c]{11}}
\put(1,2){\makebox[0pt][c]{2}}
\put(2,2){\makebox[0pt][c]{6}}
\put(3,2){\makebox[0pt][c]{10}}
\put(4,2){\makebox[0pt][c]{12}}
\put(3,3){\makebox[0pt][c]{1}}
\put(4,3){\makebox[0pt][c]{4}}
\put(4,4){\makebox[0pt][c]{3}}
\put(5,4){\makebox[0pt][c]{5}}
\put(6,4){\makebox[0pt][c]{8}}
\end{picture}
\]
Of course, given a whatever ordering of the vertices in $V(G)$, maps
$\sigma$ are naturally identified with permutations in
$\mathfrak{S}_N$, and the fraction of maps satisfying all the
constraints is given by
\begin{equation}
Z_G = \frac{1}{N!}
\, \#
\big\{ 
\sigma \in \mathfrak{S}_N
\,:\,
\forall 
(ij) \in \vec{E}, 
\;
\sigma(i)>\sigma(j)
\big\}
\ef.
\end{equation}
There are many other applications in Combinatorics and Statistical
mechanics.  A connection is with the counting of acyclic orientations
of a graph, a special evaluation of its Tutte polynomial, which is a
worst-case \#P problem (see e.g.~\cite{BL}).

For a given (unoriented) graph $G=(V,E)$, an acyclic orientation
$\phi$ is a choice of orientation for the edges, such that no oriented
cycles are created. Call $\mathcal{A}(G)$ the set of such $\phi$'s,
and, for any $\phi$, call $G(\phi)$ the corresponding oriented acyclic
graph. Any non-degenerate function $\tau$ from $V$ to a totally
ordered set induces an acyclic orientation: just orient the edge
$(ij)$ from $i$ to $j$ if $\tau(i)>\tau(j)$. Call $G[\tau]$ the
induced oriented acyclic graph.  Given a whatever ordering of the
vertices, permutations are special cases of
valid functions~$\tau$.

The uniform measure over $\mathfrak{S}_N$ is easy to study
analytically, or to exactly sample.  However, it induces a biased
measure over $\mathcal{A}(G)$ (through the natural function
$\phi(\tau)$ as the orientation $\phi$ such that $G[\tau] \equiv
G(\phi)$).  The corresponding bias factor is exactly
$Z_{G[\tau]}^{-1}$, so that, for example,
\be
\frac{1}{N!}
\sum_{\tau \in \mathfrak{S}_N}
Z_{G[\tau]}^k
=
\sum_{\phi \in \mathcal{A}(G)}
Z_{G(\phi)}^{k+1}
\ef,
\ee
and in particular the average of $Z_{G[\tau]}^{-1}$
is related to the cardinality of~$\mathcal{A}(G)$.

\begin{figure}[b]
\setlength{\unitlength}{24pt}
\begin{picture}(9,6)
\put(0,0){\includegraphics[width=216pt]{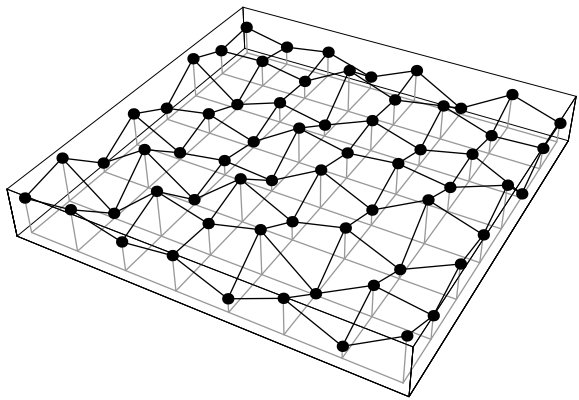}}
\end{picture}

\caption{\label{fig.excont}An example of corrugated surface in a 
  $8 \times 8$ square geometry, and ``continuous'' formulation as
  in equation~(\ref{eq.contf}).}
\end{figure}

Label the vertices with indices from $1$ to $N$.
An equivalent formulation of the problem is to ask for the Lesbesgue
measure, in the interval $[0,1]^N$, of the vectors 
$x = \{ x_i \}_{1\leq i \leq N}$
such that $x_i \geq x_j$ for each oriented edge $(ij)$ in the graph,
oriented from $i$ to $j$.
Indeed, at the aims of $N$-dimensional measure, we can neglect
configurations with repeated entries, and the constraint only depends
on the ordering of the variables, so that
\begin{equation}
\label{eq.contf}
Z_G = 
\int_{[0,1]^N} 
\!\!\!\!\!\!
\mathrm{d}x_1 \cdots \mathrm{d}x_N
\!\!
\prod_{(ij) \in \vec{E}(G)}
\!\!\!
\theta(x_i - x_j)
\ef,
\end{equation}
(here $\theta(x) = 1$ if $x \geq 0$ and $\theta(x) = 0$ if $x<0$). 
This alternate perspective gives justice to the name of 
``Corrugated Surfaces'' for the configurations on the grid (see Figure~\ref{fig.excont}), and makes
explicit another specialization of the generalized model, the
``Bead Model"~\cite{Boutillier}, corresponding to a realization on a
square lattice rotated by an angle $\pi/4$, and the edges being
directed, say, in the down-right and down-left directions.

The more general problem of understanding the statistics of local minima in a random landscape defined on a graph, which arises in the study of a point particle in a random potential or, for example, by looking at the energy landscape corresponding to configurations in many-body systems, was addressed in~\cite{Nechaev,MajMart}.
In particular in~\cite{MajMart}, instead of looking at typical configurations, the interest is shifted to the probability of large deviations, that is of configurations which are far from typical, as the configurations in which the local minima are maximally packed, i.e. Corrugated Surfaces. They provide both analytical (for Cayley trees) as numerical estimates (for hypercubic lattices) for the constant $\gamma$ which controls  the number of those configurations in the limit of large $N$, as $Z_G \sim \gamma^{-N}$. But the numerical results don't rely on Monte Carlo simulations.
 
In this paper we present a Monte Carlo algorithm to ``exactly", that is without any bias, sample
configurations $x \in [0,1]^N$ satisfying our constraints, for an
arbitrary given oriented acyclic graph $G$.


The algorithm runs under the paradigm of Wilson and Propp ``Coupling
From The Past'' (CFTP) \cite{WP1}, in a variant which allows for
continuous-valued variables (the original CFTP setting is described
for a discrete configuration space). As an application, we calculate
numerically the large-size asymptotics (for $N \to \infty$) of the
number of corrugated surfaces in two dimensions, with four digits of
precision, under some hypothesis on the scaling exponents of the
finite-size corrections (that we motivate theoretically in the
following).

At the light of this observation, it is not surprising that Corrugated
Surfaces and the Bead Model allow for exact sampling using CFTP, as
the relative discretized variants, hard-core lattice gas on the square
lattice and lozenge tilings, are among the most studied applications
of discrete CFTP or similar techniques (see \cite{RandRec} for the
hard-core gas, plus \cite{Wreadonce} for a discrete-variable
continuous-space version, and \cite{LozW} for Lozenge tilings).

The extension of the method to continuous variables is not a big deal,
and we do not claim much originality in this.  However it is not either
obvious \emph{a priori}, as a crucial point is the possibility that
two coupled random Markov chains reach coalescence, an event that could
na\"ively be though to have zero probability in Lesbesgue sense. Some
of these aspects have been already considered in the literature
\cite{Mitha}, while issues of different sort, and more specific to our
problem, are discussed in detail in the following.

\section{Coupling from the Past}
\label{sec.cftp}

\noindent
As we said in the introduction, we used an ``exact sampling''
algorithm, i.e.~an algorithm that allows to extract, in a finite time,
a configuration in an ensemble with a given measure, without any bias.
For a general discussion on Monte Carlo methods in Statistical Mechanics we refer to~\cite{Sokal}
where both the initialization bias, that is the source of {\em systematic error} related to ``thermalization", which is controlled by the {\em exponential} autocorrelation time, and the {\em statistical error}, controlled by the {\em integrated} autocorrelation time at ``equilibrium" are fully discussed.
See also~\cite{CPS} for an example of rigourous bounds on the autocorrelation times related to physical observables for Metropolis algorithms.


The \emph{Coupling From The Past} method (CFTP) also makes use of 
Markov Chain processes. However, the two fundamental concepts of
evolving a set of coupled chains, and analysing the dynamics backward
in time, allow to determine a paradigm which leads to perfectly
unbiased samples \cite{WP1}.

A coupled Markov chain is a process involving a set of configurations
(say, $k$ of them), which evolve under a dynamics being a valid
(ergodic equilibrium) dynamics for each chain, but using the very same
sequence of randomly-generated numbers in the different copies. This
makes it a non-ergodic dynamics for the $k$-uple as a whole. In
particular, if any two configurations become the same at some time
$t$, they will not anymore evolve into distinct configurations. The
first time at which all $k$ chains reach the same configuration
is called \emph{coalescence time}, and the coupled chain is said to
have \emph{coalesced} in this case.

Consider a conceptual experiment in which we follow the evolution of
all the allowed configurations, with the chains weighted according to
the measure on the ensemble. We have that, at least at the initial
time, averages on the $k$ copies of the chain correspond to
statistical averages in the ensemble. Furthermore, this property must
be preserved by the time evolution, and in particular must be true
also in a limit $t \to \infty$, in which, if the average time of
coalescence is finite, we have a single configuration.

In order not to introduce a bias due to the ``observation'' that the
chains have coalesced, one must use a protocol of ``looking backward''
in time, i.e.~imagine to perform the coupled-chain experiment above
starting at a sequence of negative times, for example $t_g=-2^g$, and
stopping the dynamics at $t=0$. Each finite state of the coupled chain
has the property that averaging on all the survived different copies,
with the measure induced by the (averaged) time evolution, would
provide the appropriate statistical averages, and in particular, at
the first \emph{generation} $g$ such that the chain has coalesced, the
only survived final state is exactly sampled.

Following the full set of possible states is clearly
unfeasible. However, there is a case in which one can certificate that
all the states in the conceptual experiment above have coalesced, by
just following a coupled Markov Chain with $k$ of order 1. This
happens if the space of configurations has a structure of
(mathematical) lattice,
i.e.~there is a structure of partial ordering $\preceq$,
and two special states $O$ and $I$ such that, for each $X$ in the
lattice,
$O \preceq X \preceq I$~\cite{birk}. Furthermore, it is required that
the (ergodic reversible) Markov Chain dynamics, whose equilibrium
distribution is the desired measure on the ensemble, preserves the
ordering, i.e., the coupled evolution of two ordered configurations
$X_t \preceq Y_t$ leads to ordered configurations $X_{t+1} \preceq
Y_{t+1}$.

In this case, it suffices to work with $k=2$, and initial states $O$
and $I$, in order to have that if $O$ and $I$ have coalesced, also all
configurations in between did it.

In summary, we can extract a general ``CFTP protocol'' for sampling on
a lattice. Given an ensemble, suppose that you can find a lattice
structure on the space of configurations, and an ergodic reversible
dynamics which preserves the ordering. Then, iteratively for
$g=0,1,2,\ldots$, run the $k=2$ coupled Markov Chain with initial
states $(O,I)$. It is important that, in the last $2^g$ times of the
simulation at generation $g+1$ you use the same set of random numbers
used in all the $2^g$ times of the run at generation $g$.  Stop the
simulation at the first generation $g_{\star}$ such that the chain has
reached coalescence: the exactly sampled configuration is the state at
time~$2^{g_{\star}}$.



\section{A CFTP algorithm for our model}
\label{sec.alg}

\noindent
Given our directed acyclic graph $G=(V,\vec{E})$,
for each vertex $i$ we define $\mathcal{N}_{+}(i)$ as the set of
vertices $j$ such that $(ji) \in \vec{E}(G)$, and $\mathcal{N}_{-}(i)$
as the set of vertices $j$ such that $(ij) \in \vec{E}(G)$. Then, a
restatement of the constraint $\prod_{(ij)} \theta(x_i - x_j)$
is that, for each vertex $i \in V(G)$,
\begin{equation} 
\max_{j \in \mathcal{N}_{-}(i) } \big( x_j \big)
\leq
x_i
\leq 
\min_{j \in \mathcal{N}_{+}(i) } \big( x_j \big)
\end{equation}
It is easy to devise an ergodic Monte Carlo chain with single-variable
heat-bath moves.  More explicitly, at each time step
\begin{enumerate}
\item Choose $i$ at random uniformly in $V(G)$;
\item Choose $z$ at random uniformly in $[0,1]$;
\item Replace $x_i$ by $z$ if it happens that
\begin{equation*}
\max_{j \in \mathcal{N}_{-}(i) } \big( x_j \big)
\leq
z
\leq
\min_{j \in \mathcal{N}_{+}(i) } \big( x_j \big)
\ef.
\end{equation*}
\end{enumerate}
Remark how conversely, in the discrete formulation, such a dynamic
would have been ``quenched'' by the further, highly non-local
constraint that all $\sigma(i)$'s are distinct.

The simple Monte Carlo chain above has the remarkable property of
being suitable for CFTP, in the way described in Section
\ref{sec.cftp}, this being another advantage of the continuous
formulation, w.r.t.~the one in terms of permutations.  Consider the
space $S \subseteq [0,1]^N$ of valid vectors $x$, under the natural
partial ordering $x \preceq y$ if $x_i \leq y_i$ for all $i \in
V(G)$. For any graph $G$, this space is a Lattice, as there are both a
$O$ and a $I$ element, corresponding to the vectors $\vec{0}$ and
$\vec{1}$ respectively.

Then, the second condition for CFTP is that, given two configurations
$x$ and $y$ at time $t$ such that $x \preceq y$, the coupled time
evolution preserves the ordering. This is easily seen, with the help
of the crucial observation that, for any vertex $i$, with the
definitions
\begin{align}
j' 
&= 
\argmin_{\mathcal{N}_{+}(i) }(y_j)
\ef;
&
j''
&=
\argmax_{\mathcal{N}_{-}(i) }(x_j)
\ef;
\end{align}
one has
\begin{subequations}
\label{eq.4357}
\begin{align}
\min_{j \in \mathcal{N}_{+}(i) } \big( x_j \big)
\leq 
x_{j'}
& \leq
y_{j'}
=
\min_{j \in \mathcal{N}_{+}(i) } \big( y_j \big)
\ef;
\\
\max_{j \in \mathcal{N}_{-}(i) } \big( x_j \big)
=
x_{j''}
& \leq 
y_{j''}
\leq
\max_{j \in \mathcal{N}_{-}(i) } \big( y_j \big)
\ef.
\end{align}
\end{subequations}
Call
\begin{align}
x_+ 
&= 
\min_{j \in \mathcal{N}_{+}(i) } \big( x_j \big)
\ef;
&
x_-
&=
\max_{j \in \mathcal{N}_{-}(i) } \big( x_j \big)
\ef;
\end{align}
The range of values in which the new candidate variable
$z$ is accepted in $x_i$ is the interval $[x_-, x_+]$, and similarly
for $y$.
The statement in (\ref{eq.4357}) is that $x_{\pm} \leq y_{\pm}$ if $x
\preceq y$.   If $x_+ > y_-$, there is a probability
$
(x_+ - y_-)/(y_+ - x_-)
$
that, given that a move occurs, the number of indices $i$ for which
$x_i \equiv y_i$ increases by one.
If instead $x_i \equiv y_i$ at some time (and in this case it must be
$x_+ > y_-$), there is a probability
$
1 - (x_+ - y_-)/(y_+ - x_-)
$
that this number
decreases by one. Calling $C(x,y) = \{ i: x_i \equiv y_i \}$, we can
define a distance parameter $d(x,y) = |C(x,y)|$, and the reasonings
above describe a non-trivial hopping dynamics in the parameter $d(x,y)
\in \{0, \ldots, N\}$.  The chain starts from $d(x,y)=N$ and reaches
coalescence when $d(x,y)=0$, which is thus a fixed point of the
induced restricted dynamics.

One could na\"ively think that the extrema $d=0$ and $d=N$ are
strongly repulsive, in a way that becomes stronger with lattice size,
because of ``entropic reasons'' (e.g., for $d=0$, if $|C| =
\mathcal{O}(N)$ and $|V(G) \setminus C| = \mathcal{O}(1)$, we have a
relative factor $1/N$ for choosing an index in the second set). At
least in graphs with low degree, say bounded by $k$, this is the case
for the $d=N$, but \emph{not} for $d=0$. Indeed, if the dynamics
chooses a site in $C$ such that all its neighbours are in $C$, the
distance can not increase at that time step, so that the ratio between
the two rates $d \to d \pm 1$ is bounded by $k$. The underlying
mechanism is somehow related to the fact that na\"ive entropic reasonings
hold for equilibrium statistical mechanics, but require more care in
non-ergodic dynamics such as in the coupled Markov Chain.

\section{From Exact Sampling to the calculation of the Free Energy}
\label{sec.FE}

\noindent
It is well known \cite{JVV} (see also \cite[sec.~3.2]{MJbook}) that the problems of
(approximated) counting  and  of uniform sampling
are closely related. In particular, there is a way of calculating
numerically the free energy of a model of interest if we can
generalize the model introducing an extra parameter $\epsilon$ such
that
\begin{itemize}
\item at $\epsilon=1$ we recover the model of interest;
\item at $\epsilon=0$ the free energy is known exactly;
\item exact sampling is available in the range $\epsilon \in [0,1]$;
\item the (unnormalized) Gibbs measures $\mu_{\epsilon}$, for
  different values of $\epsilon$, are absolutely continuous (in either
  direction).
\end{itemize}
Actually, instead of the last point, it is sufficient to have the
weaker but more technical statement of having a Radon-Nikodym
derivative of a measure at a value $\epsilon$ w.r.t.~a measure at a
value $\epsilon'$, for a suitable set of pairs $(\epsilon, \epsilon')$
(see \cite{AGS} for the pertinent definitions). Indeed, if, for
example, for $\epsilon < \epsilon'$ we have $\mu_{\epsilon} \ll
\mu_{\epsilon'}$, then we have a function $g_{\epsilon, \epsilon'}(x)$
such that
\be
|\mu_{\epsilon}| 
:= \int
\mathrm{d}\mu_{\epsilon}(x)
= \int
g_{\epsilon, \epsilon'}(x)\,
\mathrm{d}\mu_{\epsilon'}(x)
\ee
and thus, by defining the free energy as 
$F(\epsilon)=\ln |\mu_{\epsilon}|$ (we neglect overall constants
customary in thermodynamics), we have
\be
F(\epsilon) = F(\epsilon') 
+ \ln \eval{g_{\epsilon, \epsilon'}}_{\epsilon'}
\ef,
\ee
which can be used telescopically in order to obtain $F(1)$ from $F(0)$
and the evaluation of $\ln \eval{g_{\epsilon_i, \epsilon_{i+1}}}$
through the exact-sampling algorithm run at some sequence of
$\epsilon_i$'s, sufficiently dense that the statistics of 
$g_{\epsilon_i, \epsilon_{i+1}}$ is significative.
Also, if one can define
\be
\label{eq.defgt}
\tilde{g}(\epsilon)
=
\lim_{\delta \to 0}
\frac{\ln \eval{g_{\epsilon + \delta, \epsilon}}_{\epsilon}
 }{\delta}
\ef,
\ee
and this function is smooth in $\epsilon$,
the free energy for the model of interest would be given by
\be
F(1) = F(0) + \int_0^1 \!\!
\mathrm{d}\epsilon \, \tilde{g}(\epsilon)
\ef,
\ee
and it could be more efficient to find a reasonable continuous fit of
$\tilde{g}(\epsilon)$ from simulations run at some sequence of
$\epsilon_i$'s.

In our case of corrugated surfaces, for a configuration $x$ define
$W[x] \subset V(G)$ as the set of vertices being maxima and in the
range $[0, \smfrac{1}{2} ]$, union the ones being minima and in the
range $[\smfrac{1}{2}, 1]$.  Call $m(x)=|W[x]|$. Then choose
\be
\label{eq.muepsCS}
\mu_{\epsilon}(x) = 
\epsilon^{m(x)}
\prod_{(ij) \in \vec{E}} \theta(x_i-x_j)
\ef.
\ee
It is evident that $\epsilon=1$ corresponds to our model, and that at
$\epsilon=0$ we just have $Z_G=2^{-N}$. 

The Markov Chain introduced in Section \ref{sec.alg} is easily
generalized to the introduction of the parameter $\epsilon$. Just the
second point, where we ask to extract $z$ uniformly in the interval
$[0,1]$, has to be replaced by the measure on $[0,1]$
\[
p_{\epsilon}^{(\pm)}(z) = 
\frac{2}{1+ \epsilon}
\Big( \epsilon + 
(1-\epsilon) \, \theta \big( \pm \big(z - \smfrac{1}{2} \big) \big) 
\Big)
\]
with $+/-$ if we are performing the move respectively on the position
of a maximum or a minimum. The plot of $p_{\epsilon}^{(\pm)}(z)$ is just
\[
\setlength{\unitlength}{16pt}
\begin{picture}(10.5,3.8)(0,0.5)
\put(0,0.15){\includegraphics[scale=0.8, bb=0 220 210 310,
    clip=true]{fig_grid1.eps}}
\put(0.2,0.2){$\scriptstyle 0$}
\put(1.6,0.2){$\scriptstyle 1/2$}
\put(3.4,0.2){$\scriptstyle 1$}
\put(0.,2.75){$\scriptstyle 1$}
\put(-0.45,1.95){$\scriptstyle \frac{2 \epsilon}{1+\epsilon}$}
\put(-0.45,3.5){$\scriptstyle \frac{2}{1+\epsilon}$}

\put(6.2,0.2){$\scriptstyle 0$}
\put(7.6,0.2){$\scriptstyle 1/2$}
\put(9.4,0.2){$\scriptstyle 1$}
\put(6.,2.75){$\scriptstyle 1$}
\put(5.55,1.95){$\scriptstyle \frac{2 \epsilon}{1+\epsilon}$}
\put(5.55,3.5){$\scriptstyle \frac{2}{1+\epsilon}$}
\end{picture}
\]
Specialization of the quantity in (\ref{eq.defgt}) is easily achieved:
\be
g_{\epsilon, \epsilon'}(x)
=
\left( \frac{\epsilon}{\epsilon'} \right)^{m(x)}
\ee
and, taking the limit,
\be
\tilde{g}(\epsilon) =
\frac{1}{\epsilon} \eval{m(x)}_{\epsilon}
\ef.
\ee
The apparent singularity at $\epsilon=0$ is not there, as 
$\eval{m(x)}_{\epsilon}$ vanishes linearly with $\epsilon$. Also, the
potential risk of having an explosion of data noise at $\epsilon=0$ is
easily avoided: as this point is the trivial limit of the theory, it
is easy to match the data near $\epsilon=0$ with the first few terms
of a Cluster Expansion. Call $G(L)$ the graph corresponding to a grid
of side $L$, and define the intensive free energy as
\be
f(\epsilon;L)
=
\frac{1}{L^2} \ln Z_{G(L)}
\ef,
\ee
then the Cluster Expansion, in powers of $\epsilon$ and inverse powers
of $L$, gives
\be
\label{eq.cluster}
f(\epsilon;L)
=
-\ln 2
+ \frac{\epsilon}{5}
- \left( \frac{4 \, \epsilon}{15} \right)^2
\!\!
+ \frac{\epsilon}{5 L}
+ \mathcal{O}
\!\left(\epsilon^3, \smfrac{\epsilon^2}{L}, \smfrac{\epsilon}{L^2}
\right)
\ef.
\ee
The observable $W[x]$ has also an appealing interpretation. It is
clear that no adjacent vertices can be simultaneously in $W$, so, for
each configuration $x$, $W$ is a hard-core gas configuration (or an
\emph{independent set}). The parameter $\epsilon$ plays the role of an
effective fugacity in the gas, and, although the correspondence is not
perfect,
we can imagine
that, if any criticality at all appears in the model, it will be in the
same universality class of the one of the two-dimensional hard-core
lattice gas. 

For the general case, one can alternatively use the intuitive
``temperature'' parametrization. Allow for all configurations 
$x \in [0,1]^N$, and define $h(x)$ as the number of edges whose
constraint is not satisfied. Then we can write
\be
\mu_{\epsilon}(x) = 
\prod_{(ij) \in \vec{E}} 
\big( \epsilon \,
\theta(x_i-x_j)
+
(1-\epsilon) \big)
\ef.
\ee
Again $\epsilon=1$ corresponds to our model, while at
$\epsilon=0$ we just have $Z_G=1$. Similarly, 
\be
g_{\epsilon, \epsilon'}(x)
=
\left( \frac{1-\epsilon}{1-\epsilon'} \right)^{h(x)}
\ee
and, taking the limit,
\be
\tilde{g}(\epsilon) =
- \frac{1}{1 - \epsilon} \eval{h(x)}_{\epsilon}
\ef.
\ee
Again the apparent singularity at $\epsilon=1$ is not there, as 
$\eval{h(x)}_{\epsilon}$ vanishes linearly at that point. However, in
this case the increase of noise at $\epsilon=1$ is unavoidable, this
being the reason why we have chosen the \emph{ad hoc} parametrization
(\ref{eq.muepsCS}) for our numerical simulations of corrugated
surfaces.

\begin{figure}
\setlength{\unitlength}{24pt}
\begin{picture}(9,5.7)(-0.2,0)
\put(0,0.2){\includegraphics[width=216pt]{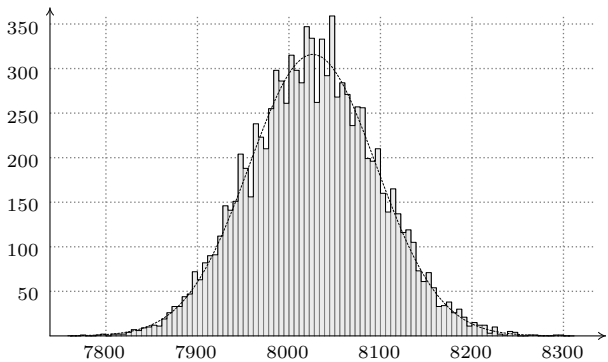}}
\put(0.7,0){\scriptsize 7800}
\put(2.146,0){\scriptsize 7900}
\put(3.592,0){\scriptsize 8000}
\put(5.038,0){\scriptsize 8100}
\put(6.484,0){\scriptsize 8200}
\put(7.93,0){\scriptsize 8300}
\put(0,0.9){\makebox[0pt][r]{\scriptsize 50}}
\put(0,1.6){\makebox[0pt][r]{\scriptsize 100}}
\put(0,2.3){\makebox[0pt][r]{\scriptsize 150}}
\put(0,3.0){\makebox[0pt][r]{\scriptsize 200}}
\put(0,3.7){\makebox[0pt][r]{\scriptsize 250}}
\put(0,4.4){\makebox[0pt][r]{\scriptsize 300}}
\put(0,5.1){\makebox[0pt][r]{\scriptsize 350}}
\end{picture}

\caption{\label{fig.gau}Distribution of $L^2 m(x)$, averaged over
  $10^4$ exactly sampled configurations, on a system with size
  $L=256$, at $\epsilon=1$. The fit is with a Gaussian.}
\end{figure}

\section{Numerical data for Corrugated Surfaces}
\label{sec.data}

\noindent
In order to obtain a reliable estimate we performed our simulations for values of $L= 16$, $32$, \ldots, $256$,
and at 20 different values of $\epsilon$, equally spaced, in the
relevant interval, in the parametrization $\epsilon/(1+\epsilon)$.  For
each value of $L$ and $\epsilon$, we performed $10^4$ independent
runs, providing us a set of numerical values for the quantities
\be
\label{eq.minfL}
\frac{\partial}{\partial \epsilon}
f(\epsilon;L)
=
\frac{\eval{m(x)}_{\epsilon}}
{L^2 \epsilon}
\ef.
\ee
For each value of $\epsilon$ in our analysis, the distribution of
$m(x)$ in the output configurations is well-fitted by a Gaussian
(cfr.~for example Figure~\ref{fig.gau}),
in agreement with the fact that, if $m(x)$ is a good order
parameter, as we expect from the analogy with the hard-core lattice
gas, we are in a regime with a single Gibbs phase. 
We checked that our numerical implementation provides exactly the results on the non-trivial case $L=3$ which can be computed analytically.
For the larger values of $L$, at each value of
$\epsilon$, the numerical values of 
$\frac{\partial}{\partial \epsilon} f(\epsilon;L)$ are in good
agreement with a finite-size description in which the first correction
scales with $1/L$. So, in a fit of the form
\be
\frac{\partial}{\partial \epsilon}
f(\epsilon;L) 
\sim 
\frac{\partial}{\partial \epsilon}
f(\epsilon) 
+ \frac{A(\epsilon)}{L}
+ \frac{B(\epsilon)}{L^2}
\ef,
\ee
we extrapolated the asymptotic values
\be
\label{eq.minf}
\frac{\partial}{\partial \epsilon}
f(\epsilon)
=
\lim_{L \to \infty} 
\frac{\eval{m(\epsilon)}_{\epsilon}}
{L^2 \epsilon}
\ef.
\ee
\begin{figure}
\setlength{\unitlength}{24pt}
\begin{picture}(9,6.6)
\put(0,0){\includegraphics[width=216pt]{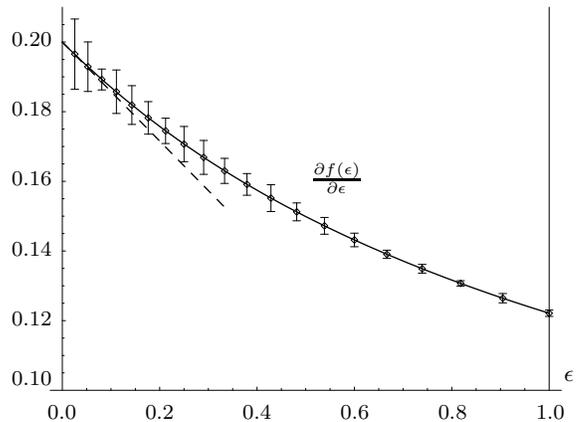}}
\put(5,3.8){$\frac{\partial f(\epsilon)}{\partial \epsilon}$}
\put(0.86,0.12){{\footnotesize 0.0}}
\put(2.40,0.12){{\footnotesize 0.2}}
\put(3.94,0.12){{\footnotesize 0.4}}
\put(5.48,0.12){{\footnotesize 0.6}}
\put(7.02,0.12){{\footnotesize 0.8}}
\put(8.56,0.12){{\footnotesize 1.0}}
\put(9,0.72){$\epsilon$}
\put(0.3,0.65){{\footnotesize 0.10}}
\put(0.3,1.6){{\footnotesize 0.12}}
\put(0.3,2.7){{\footnotesize 0.14}}
\put(0.3,3.8){{\footnotesize 0.16}}
\put(0.3,4.9){{\footnotesize 0.18}}
\put(0.3,6.0){{\footnotesize 0.20}}
\end{picture}

\caption{\label{fig.gee}Numerical data for the function in
  (\ref{eq.minf}). The natural error bars are invisible. In the
  figure, we magnified the errors by a factor 100 in order to
  highlight the different relative error in the two regimes of
  $\epsilon \to 0$ and $\epsilon = \mathcal{O}(1)$. The linear 
  behaviour at $\epsilon = 0$ is
  deduced from the Cluster Expansion in~(\ref{eq.cluster}).}
\end{figure}
A scrupolous statistical analysis of the errors, just within the
(well-verified) assumption on the exponent of the leading finite-size
correction, leads to the results in Figure~\ref{fig.gee}. Finally,
numerical integration of a polynomial interpolation of the data (with
a polynomial of degree 5, determined by analysis of structure in the
errors) provided us with the result
\be
f = -0.53967 \pm 0.00054
\ef.
\ee
The parameter $\gamma$  is obtained from
$\gamma = \exp (- f)$. 
Our numerical analysis on the square lattice gives
\be
\gamma = 1.7154 \pm 0.0009
\ee
with purely statistical errors.
This must be compared with the previous numerical estimate~\cite{MajMart}
\be
\gamma_{\rm MM}(2)= 1.6577 \pm 0.0006
\ef,
\ee
which is definitely out of the estimated errors. Even if the difference appears to be small it is interesting to observe that,
as the estimate in~\cite{MajMart} for the three-dimensional cubic lattice is
\be
\gamma_{\rm MM}(3)= 1.7152 \pm 0.0010
\ee
the discrepancy is of the same size of the estimated difference between the two and three dimensional constants.

\section{Analysis of the times to coalescence}
\label{sec.times}

\begin{figure}
\setlength{\unitlength}{24pt}
\raisebox{-66pt}{
\begin{picture}(7.1,5.3)(-0.1,0)
\put(0.2,1.25){{\footnotesize 0.2}}
\put(0.2,2.175){{\footnotesize 0.4}}
\put(0.2,3.10){{\footnotesize 0.6}}
\put(0.2,4.025){{\footnotesize 0.8}}
\put(0.2,4.95){{\footnotesize 1.0}}
\put(-0.35,4.5){$p_L(g)$}
\put(1.730,0.1){{\footnotesize 20}}
\put(2.863,0.1){{\footnotesize 40}}
\put(3.996,0.1){{\footnotesize 60}}
\put(5.129,0.1){{\footnotesize 80}}
\put(6.17,0.1){{\footnotesize 100}}
\put(6.9,0.05){$L$}
\put(1.3,5.25){{\footnotesize $g:$}}
\put(1.9,5.25){{\footnotesize 14}}
\put(2.45,5.25){{\footnotesize 15}}
\put(3.1,5.25){{\footnotesize 16}}
\put(4.0,5.25){{\footnotesize 17}}
\put(5.2,5.25){{\footnotesize 18}}
\put(6.6,5.25){{\footnotesize 19}}
\put(0,0){\includegraphics[width=168pt]{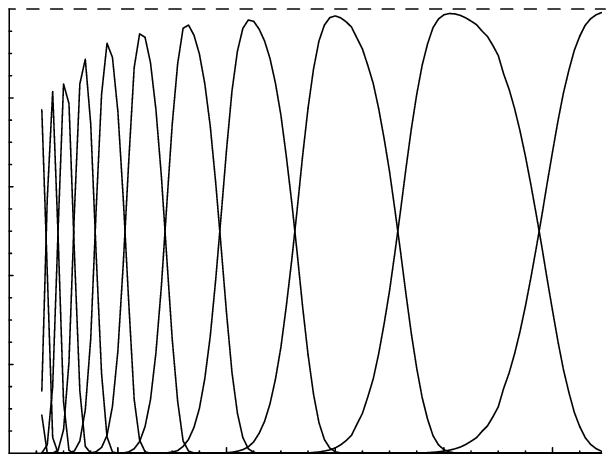}}
\end{picture}
}
\qquad
\begin{tabular}{|c|c|}
\hline
 $g$ & $L(g)$  \\
\hline
\vspace*{-2pt}%
{\footnotesize  9} & {\footnotesize 6.802} \\
\vspace*{-2pt}%
{\footnotesize 10} & {\footnotesize 8.943} \\
\vspace*{-2pt}%
{\footnotesize 11} & {\footnotesize 11.85} \\
\vspace*{-2pt}%
{\footnotesize 12} & {\footnotesize 15.82} \\
\vspace*{-2pt}%
{\footnotesize 13} & {\footnotesize 21.20} \\
\vspace*{-2pt}%
{\footnotesize 14} & {\footnotesize 28.57} \\
\vspace*{-2pt}%
{\footnotesize 15} & {\footnotesize 38.65} \\
\vspace*{-2pt}%
{\footnotesize 16} & {\footnotesize 52.45} \\
\vspace*{-2pt}%
{\footnotesize 17} & {\footnotesize 71.41} \\
{\footnotesize 18} & {\footnotesize 97.39} \\
\hline
\end{tabular}

\caption{\label{fig.bellshapes}On the left, plots of $p_L(g)$ in the
  variable $L$, for $g = 8,\ldots, 19$.
  On the right, table of the values $L(g)$
  interpolated from the data. Errors are estimated to be on the last
  digit.}
\end{figure}

\noindent
According to the ``ordinary'' CFTP protocol (not the Read-Once
protocol of \cite{Wreadonce}), the running times for each size, and
each independent instance, is essentially a power of $2$, say,
$2^g$. This means that the average complexity of the algorithm is
described through some probability for the \emph{generation} parameter
$g$, at side $L$, that we call $p_L(g)$
\be
\mathrm{Time}(L) \sim \sum_g 2^g p_L(g)
\ef.
\ee
It is a tautology that this time is finite provided that $p_L(g)$ has
a Laplace Transform at the value $-\ln 2$. It comes out that, in our
range $\epsilon \in [0,1]$, the times grow monotonically with
$\epsilon$, and even at the hardest point $\epsilon=1$ the histogram
$p_L(g)$ takes its leading contribution from one or two values of
$g$. At fixed $g$, the plot of $p_L(g)$ in the variable $L$ instead
looks like a smooth bell-shaped curve in the range $[0,1]$
(cfr.~Figure~\ref{fig.bellshapes}), and it is
fairly safe to interpolate the (continuous) value $L(g)$ of $L$ at which
the curve for $g$ and the one for $g+1$ do cross. Given this regular
behaviour, the scaling of $L(g)$ with $g$ must be related to the
scaling of the complexity, in particular, if $g(L)$ is the functional
inverse of the (obviously monotone) $L(g)$, then
\be
\mathrm{Time}(L) \sim 2^{g(L)}
\ef.
\ee
The data are well fitted by a curve of the form
\be
\label{eq.37876967}
g(L) \propto L^2 \ln L
\ef,
\ee
so the complexity seems to be only logarithmically
super-linear in the number of degrees of freedom.

As we discussed in Section \ref{sec.FE}, the pictorial interpretation
of the observable $W[x]$ suggests that our model of Corrugated
Surfaces, in its continuation to arbitrary values of $\epsilon$, is in
the same universality class as the hard-core lattice gas, so it was
natural to expect that, analogously to what is proven for this model
\cite{LV, DG}, the average coalescence time is $\sim n \ln n$ for
Glauber dynamics at sufficiently small fugacities (in agreement with
(\ref{eq.37876967}) above), while it becomes worst-case hard at
sufficiently large fugacities, these values being lower- and
upper-bounds to the ``physical'' critical values $\epsilon^{\star}$
(the gap being originated by technicalities in the proof
procedure). It is not inconceivable that, for our system, all the
range $\epsilon \leq 1$ has fast coalescence times.

Furthermore, fast convergence would imply a choice of parameters which
are far from a critical point, and correspond to a single
thermodynamic phase
\cite{Weitz, WeitzT}, 
and this would imply in turns that finite-size corrections to
intensive observables scale with the ratio perimeter/area, i.e.~with
$L^{-1}$ in our two-dimensional case. This justifies the treatment of
the finite-size corrections that we have done in
Section~\ref{sec.data}.

\section{Conclusions}

\noindent
We have pointed out how the method of Coupling From The Past may be
fruitfully applied to problems with variables assuming values on a
continuum domain, and how a one-variable Heat-bath Monte Carlo chain
is suitable for CFTP in an interesting problem in this class: the
uniform sampling of height functions satisfying a set of inequalities
described by an acyclic graph.

As a specific example, we performed a numerical simulation for the
model of Corrugated Surfaces, in order to determine  the value of the free energy. 
Our precision goal was to have error bars much below the order of magnitude of the discrepancy
between  our and  the  estimate, obtained without using a Monte Carlo method, that appears in Majumdar
and Martin~\cite{MajMart}. To this aim, the use of an exact sampling
method in connection with an exact formula for the extrapolation to the thermodynamical limit
has been useful in order to rule out any possible source of systematic errors.

We have achieved our goals with a relatively small numerical effort,
because the model of Corrugated Surfaces appears to be a special value
of a one-parameter family of models, in the universality class of the
two-dimensional hard-core lattice gas, in the low-density phase.  In
this case, as expected from the literature, the mixing times are only
logarithmically super-linear, and the coalescence time for the couple
chain is of the same order of magnitude of the mixing time of the
ordinary Markov Chain.  As a final comment, we advice that, in this
regime, the strong control over the errors in the CFTP overwhelms the
small gain in terms of computational times of other Monte
Carlo algorithms.

\section*{Acknowledgements}

\noindent
We thank L.~Cantini and C.~Boutillier for fruitful discussions.


\end{document}